\documentclass[a4paper,fleqn]{cas-dc}

\usepackage[english]{babel}
\usepackage[utf8]{inputenc}
\usepackage[T1]{fontenc}
\usepackage{microtype}
\usepackage{amsmath}
\usepackage{amssymb}
\usepackage{graphicx}
\usepackage[numbers]{natbib}
\usepackage{nicefrac}
\usepackage{caption}
\usepackage{subcaption}
\usepackage{booktabs}
\usepackage{arydshln}
\usepackage{xcolor}
\usepackage[normalem]{ulem}
\usepackage{hyperref}
\usepackage[nameinlink, capitalise]{cleveref}
\usepackage[switch,columnwise]{lineno}
\usepackage{siunitx}
\DeclareSIUnit\uF{$\mu F$}

\makeatletter
\renewcommand{\fnum@figure}{Fig. \thefigure}
\makeatother

\def\tsc#1{\csdef{#1}{\textsc{\lowercase{#1}}\xspace}}
\tsc{WGM}
\tsc{QE}

\begin{document}
\let\WriteBookmarks\relax
\def\floatpagepagefraction{1}
\def\textpagefraction{.001}

\title [mode = title]{Development of the front-end electronics for a cost-effective PET-like detector system}

\author[1]{Josephine Oppotsch}[type=editor, 
							  style=chinese, 
							  orcid=0000-0003-2577-2069]
\cormark[1]
\cortext[1]{~Corresponding author. E-mail address:}
\ead{joppotsch@ep1.rub.de}
\author[1]{Nadia Böhle}[style=chinese, orcid=0009-0005-6797-1586]
\author[1]{Thomas Held}[style=chinese, orcid=0000-0001-8581-4238]
\author[1]{Mario Fink}[style=chinese]
\author[1]{Miriam Fritsch}[style=chinese, orcid=0000-0002-6463-8295]
\author[1]{Fritz-Herbert Heinsius}[style=chinese, orcid=0000-0002-9545-5117]
\author[1]{Matthias Steinke}[style=chinese, orcid=0000-0002-4153-5488]
\author[1]{Ulrich Wiedner}[style=chinese, orcid=0000-0002-9002-6583]

\affiliation[1]{organization={Ruhr-Universität Bochum},
				addressline={Institut für Experimentalphysik I}, 
				city={44801 Bochum}, 
				country={Germany}}

\begin{abstract}
Most detector systems used for positron emission particle tracking (PEPT) are very expensive due to the use of inorganic plastic scintillators combined with a high number of readout electronic channels. This work aims to reduce the overall cost of a PEPT-capable detector system by using large and cost-effective plastic scintillators and developing custom 2\,$\times$\,2 silicon photomultiplier (SiPM) arrays, preamplifiers, and discriminators. The use of long (20\,mm~$\times$~20\,mm~$\times$~1000\,mm) plastic scintillator bars read out with photodetectors only at their respective ends allows an overall smaller number of photodetectors and associated readout electronics, which in turn reduces the overall cost of the system. In addition, the development of a custom SiPM array and preamplifier allows a free selection of interconnection and readout, as most commercial producers only offer specific types of interconnections and therefore lack other connections such as serial or hybrid. Thus, several common circuit types for SiPMs and preamplifiers were tested and compared in this work, and it was found that a serial connection implemented in a hybrid interconnection for the SiPMs and an inverting preamplifier based on a high-frequency operational amplifier provided the best results for the proposed detector system. Measured with a $^{22}$Na source, the combination of SiPM array and preamplifier led to a rise time of 3.7\,ns and a signal amplitude of 175\,mV.
\end{abstract}

\begin{keywords}
\sep Plastic scintillators
\sep Positron emission particle tracking
\sep Preamplifier  
\sep Silicon photomultiplier arrays
\sep Time-of-flight
\sep \rule{4.35cm}{0.3pt}
\sep License:
\sep $\copyright$ 2023. This manuscript version is made available under the CC-BY-NC-ND 4.0 license
\sep https://creativecommons.org/licenses/by-nc-nd/4.0/ 
\end{keywords}

\maketitle

\section{Introduction}\label{introduction}
\noindent Large industrial processes have a significant influence on the environment. However, even if most people's environmental awareness has already changed and improved significantly, there are still areas where a change proves to be very difficult. Typical examples of such areas are large-scale industrial processes that deal with the processing of densely packed and moving granular material like waste incineration or the treatment of bulk solids in process, chemical, pharmaceutical, and food industry. To improve them, bulk internal information of such large, dense, and even optically opaque systems (like particle movement and gas dispersion) is needed. Positron emission particle tracking (PEPT) is able to provide some of the required information experimentally, as it can track particle motion within these systems. 

Since PEPT is based on positron-electron annihilation, this method works by placing a positron-emitting radioisotope inside the vessel under investigation. The emitted positrons form back-to-back gamma-ray pairs by their immediate (after 0.53\,mm on average or, at the latest, after 2.28\,mm \cite{positronRangeNa-22}) annihilation with an electron of the surrounding matter. The two detection points of the associated 511\,keV gamma rays can be connected by a straight line, which thus also contains their common point of origin (back-to-back gamma rays). Hence, the position of the positron-emitting radioisotope can either be determined by time-of-flight (TOF) differences of the associated gamma rays or by the intersection of at least two of these straight lines. Therefore, unlike PET, PEPT is not an imaging but a tracking method.

PEPT was developed at the University of Birmingham in the late 1980s/early 1990s \cite{PEPTDevelopment1,PEPTDevelopment2}, and has been widely used to study particle trajectories and velocities in the context of process engineering ever since. The work of Windows-Yule et al. \cite{PEPTReview} gives a comprehensive overview of the most commonly used detector systems, doping procedures, simulations, and tracking algorithms.

Most and commonly used PEPT detector systems use crystal scintillators such as sodium iodide (NaI) for photodetection. These crystal scintillators have a high light yield and good energy resolution but are very expensive and often hygroscopic. One way to reduce costs (by a factor of 80) is to switch to cost-effective plastic scintillators. Plastic scintillators are especially suitable for time measurements due to a short rise time of 0.9\,ns \cite{AnstiegszeitSzinti} but have a lower light yield than crystal scintillators (typically 25-30\,\% of NaI, \cite{SaintGobainSzintillator}), which results in a poorer energy resolution. In addition, the energy spectrum does not exhibit a photo peak, so scattered photons cannot be detected from the spectrum. Therefore, switching from crystal scintillators to plastic scintillators is not always possible. Since this study intends to track moving and densely packed particles via PEPT and TOF and thus relies on fast time resolution, high event rates, and good solid angle coverage, the poorer energy resolution, missing photo peak, and the lower light yield can be accepted. Furthermore, Monte Carlo simulations performed in previous work \cite{oppotsch2023simulation} have shown that the efficiency of the detector system, which is defined as the ratio of reconstructed decay locations to simulated decays (therefore including all individual efficiencies, such as the rate of scattered gamma rays, the detection threshold and the efficiency of the plastic scintillators) is 1.4\,\%. It could be enlarged by closing all sides (i.e., attaching scintillators above and below the vessel as well). However, this would also significantly increase the overall cost of the detector. Another less expensive way is to increase the activity of the source to obtain a higher rate of reconstructed positions. The human body can only be exposed to a certain amount of radioactivity without being permanently damaged. In the field of process engineering, there is no need to consider dose limits for patients. As a result, much higher doses can be selected if required by the accuracy of the tracking.

At this point, it is fair to state that the overall achievable spatial resolution of the intended detector system is not as good as that of a medical scanner/commonly used PEPT detector, as it is intended to develop a most cost-effective system capable of scanning large volumes as a whole with sufficient spatial resolution, but it will be able to determine the particle trajectories via both trigonometry and time-of-flight calculations. While TOF is not new in the field of medical PET, it has not been used extensively in the field of PEPT \cite{PEPTReview}.

The Jagiellonian PET (J-PET) is the first total body positron emission tomography scanner consisting of cost-effective plastic scintillators. Besides its use as a total-body scanner in medical PET \cite{J-PET_medicalPerspectives}, it also found use as the first system capable of imaging positronium and multiphotons \cite{J-PET_PositroniumImaging}. As described in \cite{J-PET_Setup}, the entire system consists of 192 scintillator bars organized in three cylindrical and concentric layers, with each bar having a dimension of 7\,mm~$\times$~19\,mm~$\times$~500\,mm. A photomultiplier is coupled to each end of the bars, and the readout is done by multi-constant-threshold boards. Thus, the design appears to be similar to the system proposed in this work (\cref{Sec:setup}). However, since the J-PET is designed as a medical PET scanner, the focus is on imaging stationary systems rather than tracking densely packed and moving granular assemblies. Furthermore, the proposed detector system is intended to be used in the field of process engineering, which means that it must be capable of detecting much larger systems than the J-PET. Hence, larger scintillator bars are required, for which a combination of silicon photomultiplier arrays (SiPM), preamplifiers, and discriminators is envisioned for photodetection and readout.

The following work addresses the development and testing of suitable SiPM arrays and preamplifiers for an envisaged cost-effective PET-like detector system capable of tracking particles in densely packed and moving granular assemblies. The proposed detector system will consist of 88 plastic scintillator bars read out from both ends by 2\,$\times$\,2 SiPM arrays. The selection and arrangement of the individual detector components, i.e., the scintillator bars and the readout electronics, not only reduces the overall cost by using plastics instead of crystals but also reduces the granularity by requiring fewer readout electronics compared to commercial PET scanners or commonly used PEPT detectors. In addition, the design of custom SiPM arrays, preamplifiers, discriminators, and the associated printed circuit boards (PCB) ensures the best possible performance of the detector system since most producers only provide certain types of connection for their arrays and readout boards, which lack connections of the SiPMs such as serial or hybrid. Further details on the design and interconnection of the SiPM arrays and preamplifiers are given in \cref{Sec:SiPM_arrays} and \cref{Sec:preamplifier}.

For the mixing and segregation of the granular assembly, a batch-operated grate system was selected. It is inspired by industrial grate firing systems (such as the ones used for the combustion of wood chips or municipal waste) but is simplified such that particle feeding and discharge are neglected. This allows for the investigation of particle transport only. It is placed in the center of the detector system, and both are described in more detail in \cref{Sec:setup}. $^{22}$Na will be used as a positron-emitting radioisotope. A first feasibility study of this combination based on Monte-Carlo simulations, as well as deeper insights into the reconstruction, can be found in \cite{oppotsch2023simulation}. The work closes with a brief conclusion (\cref{Sec:outlook}).

\hfill\\[-0.53cm]
\section{Setup} \label{Sec:setup}
\begin{figure}
 	\centering
    \includegraphics[width=0.4\textwidth]{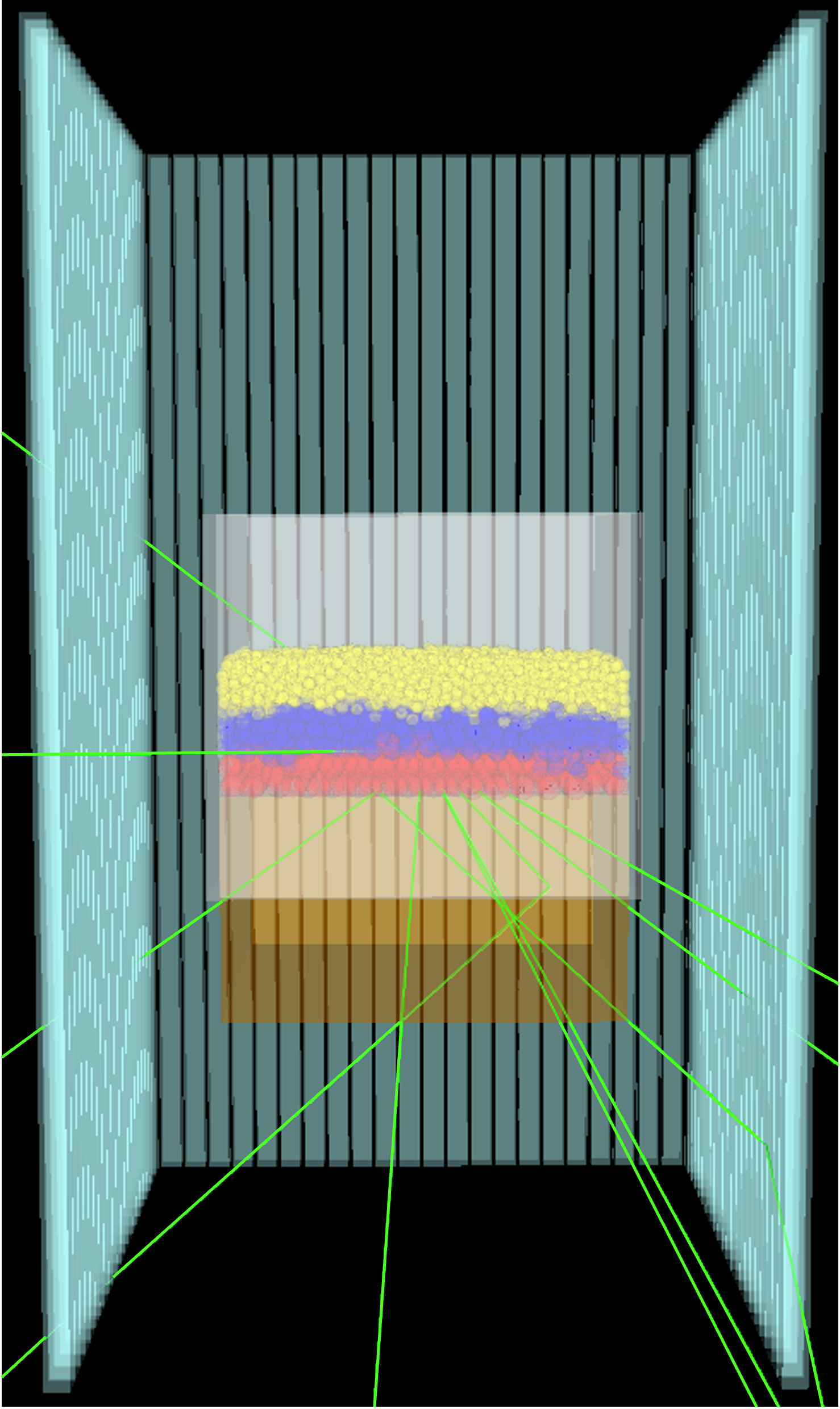}
    \caption{Simulated design of the generic grate system surrounded by the detector system. The aluminum stoking bars of the generic grate system are shown in gold and the acrylic glass box in light gray. The individual scintillator bars of the detector system are shown in light turquoise. For reasons of simplicity, the photodetectors and readout electronics are not included in this illustration. This also applies to the fourth detector wall. The green lines represent the trajectories of the gamma rays.}
    \label{Fig.1}
\end{figure}

\noindent The proposed detector system is designed to enclose the vessel under investigation from four sides (\cref{Fig.1}). Therefore, four detector walls are envisaged, consisting of 22 plastic scintillator bars each. Since the individual bars have a dimension of 20\,mm~$\times$~20\,mm~$\times$~1000\,mm, one wall will have a height of 1\,m and a length of about 0.5\,m, giving an overall detector volume of about 0.5~$\times$~0.5~$\times$~1\,m$^3$. The vessel under investigation is placed in the center of the detector system and consists (for the planned experimental measurements) of a 320\,mm~$\times$~300\,mm~$\times$~300\,mm acrylic glass box, whose bottom is formed by vertically movable stoking bars allowing the mixing of the granular assembly (\cref{Fig.1}).

\begin{figure*}
 	\centering
    \includegraphics[width=\textwidth]{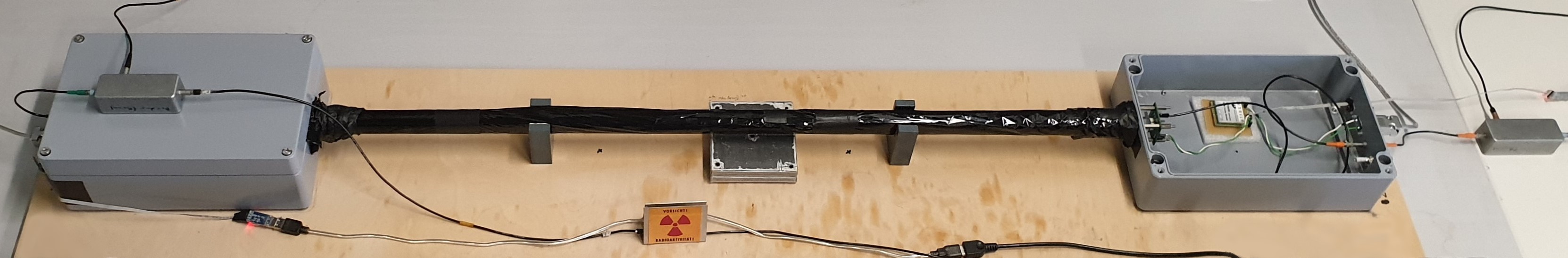}
    \caption{Image of the test setup: The grey housings are light-tight and contain the photodetectors and readout electronics. The wrapped scintillator bar can be seen in the center of the picture directly above the collimator for the $^{22}$Na source. The box on the right has been opened only for taking the image so that its content is visible. Otherwise, it is always closed, especially during measurements.}
    \label{Fig.2}
\end{figure*}

\begin{figure}
 	\centering
    \includegraphics[width=0.35\textwidth]{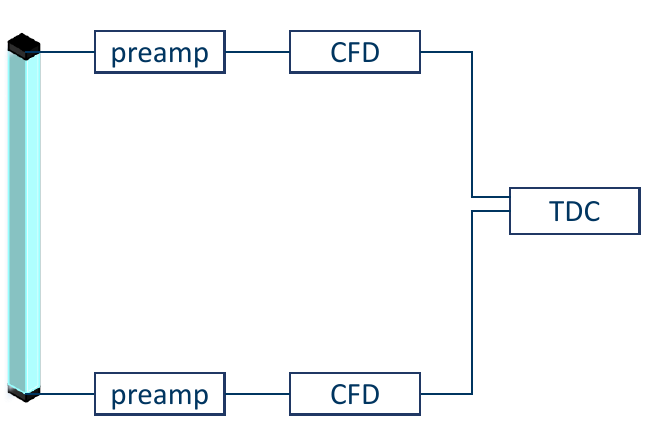}
    \caption{Sketch of the planned readout chain. The scintillator bar is shown in turquoise, and the SiPM arrays in black. ``preamp'' is short for preamplifier, ``CFD'' for constant fraction discriminator, and ``TDC'' for time-to-digital converter.}
    \label{Fig.3}
\end{figure}

Each scintillator bar is first wrapped in reflective foil and then in black foil to maximize light yield and shield from ambient light. Readout is performed by 2 $\times$ 2 silicon photomultiplier (SiPM) arrays (where one SiPM is 6\,mm $\times$ 6\,mm consisting of 22,292 35\,$\mu$m microcells), which are attached to both ends of the scintillator bars and whose signals are amplified using preamplifiers. The scintillator ends, as well as the SiPM arrays and preamplifiers, are located in light-tight boxes (\cref{Fig.2}) so that the remaining area of the scintillator ends not covered by SiPMs (256\,mm$^2$) can be left free. Triggering is performed by constant fraction discriminators (CFD), and times are measured by commercially available CAEN ``V 1190A'' multihit time-to-digital converters (TDC). A sketch of the readout chain is also shown in \cref{Fig.3}.

In order to obtain the highest possible performance and accuracy of the detector system, all components were tested for their performance, suitability, and interplay. To this end, a small test setup consisting of one scintillator bar, two light-tight aluminum housings containing the photodetection and readout electronics, and a $^{22}$Na source was built (\cref{Fig.2}). Based on the work done in \cite{oppotsch2023simulation}, which already determined the most suitable scintillator type (Saint-Gobain BC-408) and its characteristics, this work focuses on the interconnection of the SiPM arrays and preamplifiers. Therefore, the left aluminum housing (closed box in \cref{Fig.2}) always contains the previous state of the readout electronics (and will therefore be referred to as the reference setup throughout this work), and the right the one to be tested. This is done because a one-sided readout is sufficient for the performance tests of the readout electronics, and a direct comparison between the previous and the one being tested is given. Furthermore, the possibility of a direct comparison has the advantage that no radioactive source with fixed energy is needed so that the measurements can be made with cosmic rays as well.

\section{General measurement methods}
\noindent The main focus in the development of the detector is to achieve a good time resolution. For this, not only a low rise time but also a minimized jitter is important. We consider the noise being Gaussian distributed, so jitter is defined as: 
\begin{align}
   \textit{Jitter} = \frac{\sigma_n}{\nicefrac{dU}{dt}}. 
\end{align}
Here, $\sigma_n$ is the width of the Gaussian distribution, and $\nicefrac{dU}{dt}$ is the slope of the signal at trigger voltage level. To compare different circuits among each other, the amplitude (which was taken as an average over 5 maximum voltages of the waveform for the selection between the individual circuits, or 10\,-\,15 maximum voltages for the exact measurements of the final circuits) measurements were made in comparison to the amplitude $A_{ref}$ of the reference setup on the opposite side of the scintillator bar. If the noise of the SiPM circuits is the same, the jitter is in first order proportional to
\begin{align}
   J = \frac{t_{rise}\cdot A_{ref}}{A} 
\end{align}
where $t_{rise}$ denotes the rise time of the signal (defined as the time required to rise from 10\,\% to 90\,\% of the maximum voltage of the waveform) and $A$ the signal amplitude. Thus, a high amplitude is also beneficial to keep the jitter low. For this reason, the parameter $J$ was also taken into account when comparing different circuits.

For all measurements, the ``J-Series 60035'' SiPMs from onsemi were used. The supply voltage was set to $V_{Bias}$~=~29\,V for one single SiPM. This was realized with ``Caen A7585'' power supplies. The triggering was performed at the 341\,keV (or 1060\,keV for the combination of the final preamplifier and SiPM array as in \cref{Sec:dimensioningIA}) Compton edge of the $^{22}$Na source.

\section{Circuit development of a 2\,$\times$\,2 SiPM array} \label{Sec:SiPM_arrays}
\hfill\\[-1.38cm]
\subsection{Known options for SiPM interconnection and their effects}
\hfill\\[-0.65cm]
\noindent As with any electronic component, there are initially two options for interconnecting SiPMs: series interconnection and parallel interconnection. Series connection of SiPMs offers two major advantages. First, it provides ``automatic overvoltage adjustment'', which means that despite different breakdown voltages, the same overvoltage is automatically set \cite{BiasingReadout}. This is possible because in a serial connection, the same current flows through all SiPMs. Since the current-voltage characteristics for SiPMs with different breakdown voltages have the same shape except for a shift along the voltage axis, the same overvoltage is present for the same current \cite{Cryo}. Second, the total capacitance is reduced by the cells connected in series. The RC time constant of a cell, which significantly affects the rise time of a signal, is linearly dependent on the capacitance. Therefore, the smaller capacitance of series connections results in a faster rising signal. The disadvantage is that the supply voltage is proportional to the number of SiPMs connected in series, and both the amplitude and the total transported charge of the signal are reduced.

Parallel connections only reduce the amplitude while the charge is maintained (so the current peak keeps its area).\label{sec:Theopar} Another advantage of parallel connections is that the supply voltage remains the same, regardless of the number of SiPMs. Additionally, it has a better signal-to-noise ratio than the series connection. However, parallel-connected capacitors, and therefore parallel-connected SiPMs, have a larger total capacitance. This results in a higher rise time and thus poorer time resolution \cite{SiPMinNuclearPhysics}.

\begin{figure}
\centering
\includegraphics[width=0.363\textwidth]{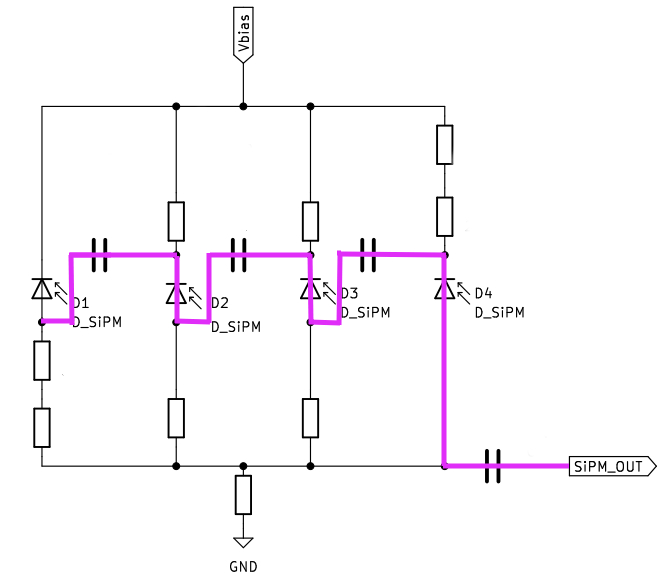}
\caption{Circuit diagram of SiPMs in a hybrid connection and its signal path (purple).}
\label{Fig.4}
\end{figure}

The hybrid configuration utilizes the advantages of both interconnections and combines them in a circuit with a parallel power supply and serial signal readout (\cref{Fig.4}). As a result, the supply voltage remains independent of the number of supplied SiPMs, and the signal rise time is minimized since all SiPMs are connected in series with coupling capacitors. The signal (marked in purple, \cref{Fig.4}) can be read out through all SiPMs in series. 
At the end, the AC signal of the SiPMs is separated from the DC component by a capacitor of a few picofarads. The bias voltage ($V_{Bias}$) is supplied in parallel with DC voltage and must correspond to $V_{Bias}$ of a single SiPM plus the voltage drop over 2R. It is also important that the resistors after the first SiPM and in front of the last SiPM are twice as large as the internal resistors \cite{Cryo}. This is to ensure that the same voltage is applied to each of the SiPMs.

\subsection{Circuit designs for interconnecting 4 SiPMs}
\begin{figure}
\centering
\includegraphics[width=0.4\textwidth]{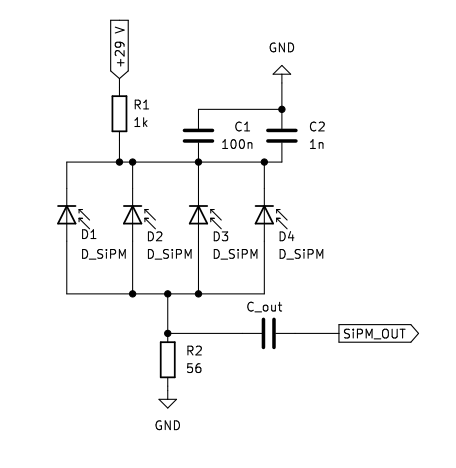}
\caption{Circuit diagram of the 4p connection.}
\label{Fig.5}
\end{figure}

\begin{figure}
\centering
\includegraphics[width=0.4\textwidth]{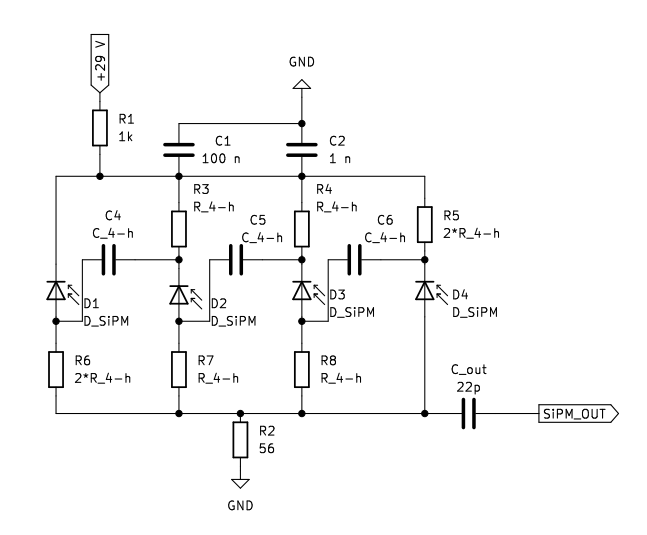}
\caption{Circuit diagram of the 4-hybrid connection.}
\hfill\\[0.1cm]
\label{Fig.6}
\end{figure}

\begin{figure}
\centering
\includegraphics[width=0.4\textwidth]{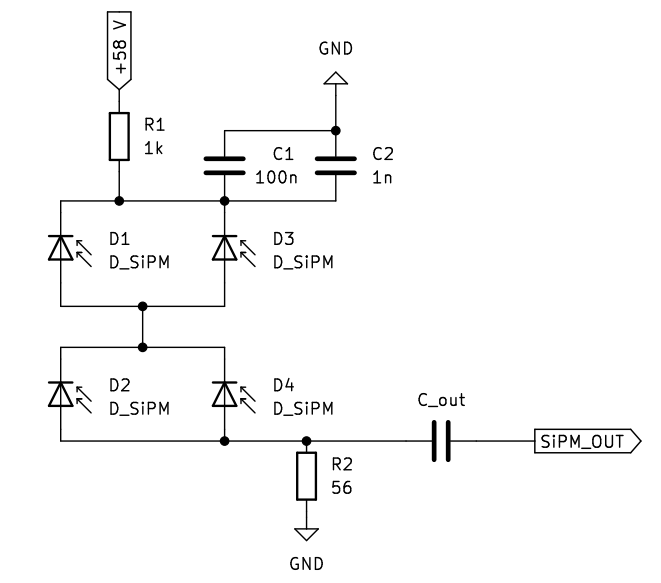}
\caption{Circuit diagram of the 2p s connection.}
\label{Fig.7}
\end{figure}

\begin{figure}
\centering
\includegraphics[width=0.35\textwidth]{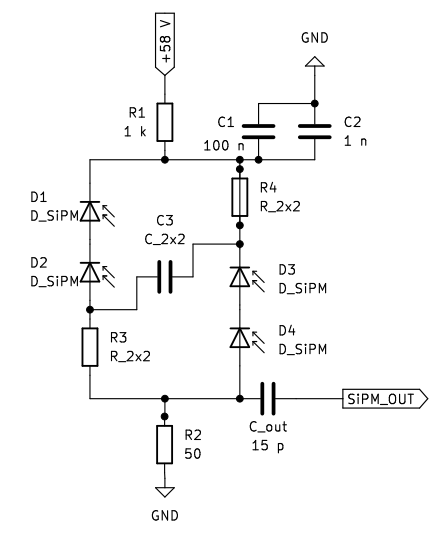}
\caption{Circuit diagram of the 2\,$\times$\,2\,-\,hybrid connection.}
\hfill\\[-0.14cm]
\label{Fig.8}
\end{figure}

\noindent The main idea is to find an interconnection scheme for SiPMs that prioritizes good time performance, considering that the amplitude can also be increased by the preamplifier later on. Therefore, a low rise time is the first priority when comparing different circuits. Since only a very small array of 2\,$\times$\,2 is required to cover the scintillator, the incorporation of serial interconnections can be seen as an option without $V_{Bias}$ becoming excessively high. The series connection exhibits steeper signal edges. Therefore, the three basic circuit types presented (serial, parallel, and hybrid connection) are first considered as options. Then, possible combinations of them are examined, leading to the following suggestions for interconnecting the four SiPMs:
\begin{itemize}
    \item 4 in series connection (4s)
    \item 4 in parallel connection (4p, \cref{Fig.5})
    \item 4 in hybrid connection (4-hybrid, \cref{Fig.6})
    \item 2 parallel connected in series connection (2p s, \cref{Fig.7})
    \item 2 serially connected in parallel connection (2s p)
    \item 2 serially connected used in a 2-stage hybrid configuration (2\,$\times$\,2-hybrid, \cref{Fig.8})
\end{itemize}
Since the power supply for the used setup has its limit at 85\,V, the option ``four SiPMs in series'' was not considered any further. The options ``2s p'' and ``2p s'' should result in the same rise time due to the same total capacitance, so just one version was tested. The idea behind a 2\,$\times$\,2-hybrid configuration (\cref{Fig.8}) is based on clear advantages in time resolution due to the series connection. 

According to the manufacturer's recommendation \cite{SiPMDatash}, capacitors are used to smooth the supply voltage. Therefore 1\,nF and 100\,nF in parallel were chosen. The resistor R2, which connects to ground, is also adopted from \cite{SiPMDatash}, and the signal is coupled via the standard output (S-out) with a 22\,pF capacitor ($C_{out}$).

\subsection{Comparison of different SiPM interconnections}
\noindent For both hybrid configurations (\cref{Fig.6} \& \cref{Fig.8}), 1\,k$\Omega$ and 2\,k$\Omega$ resistors, as well as 1\,nF capacitors, were used. The signal was decoupled with $C_{out}$~=~22\,pF and amplified through an inverting amplifier. Since only the SiPM circuits needed to be compared to each other, the amplitude was measured in comparison to the reference setup (placed on the other side of the scintillator bar). Thus, the measured values are the fractions $\nicefrac{A}{A_{ref}}$\,, which are provided in \cref{Tab.1}.

\begin{table}
\centering
\caption{Relative amplitude and rise time for different SiPM interconnections read out by a coupling capacitor ($C_{out}$ = 22\,pF) and an inverting amplifier.}
	\begin{tabular}{m{0.095\textwidth} m{0.095\textwidth} m{0.095\textwidth} m{0.095\textwidth}}
	\toprule[1pt]
	   		   	& $\nicefrac{A}{A_{ref}}$  & $t_{rise}$ (ns)  & J (ns) \\ 
	\toprule[1pt]
	4p 		   			  & 0.5 				 		& 4.7 			   & 9.4    \\
	2p s       			  & 1.7 				 		& 5.0 			   & 2.9    \\
	4-hybrid   			  & 0.7 				 		& 4.8 			   & 6.9    \\
	2\,$\times$\,2-hybrid  & 2.9 			 		& 4.3 			   & 1.5    \\
	\toprule[1pt]
	\end{tabular}
\label{Tab.1}
\end{table}

The smaller amplitude of the 4-hybrid configuration was expected. As the signal must pass through three RC high-pass filters, it becomes smaller compared to the 2\,$\times$\,2 configuration, where only one needs to be passed. It had previously been anticipated that the 2\,$\times$\,2-hybrid configuration would achieve a smaller $t_{rise}$ compared to the 4-hybrid configuration by implementing a series connection. All in all, the 2\,$\times$\,2-hybrid connection clearly shows the best performance, not only in terms of a low rise time but also in a high amplitude. Nevertheless, the 4-hybrid connection was further considered because, compared to the other circuits, the performance of hybrid ones are strongly dependent on the value of the capacitors and resistors. Therefore, a variation of those components was necessary to compare both hybrid connections. 

To cover a wide range, resistors of 1\,k$\Omega$ and 47\,$\Omega$, as well as capacitors of various magnitudes (10\,pF, 1\,nF, 100\,nF) were chosen for the 4-hybrid circuit (\cref{Tab.2}). While the rise time clearly increases with capacitance, the amplitude increases with smaller resistors and increasing capacitance. The lowest $t_{rise}$ of 3.26\,ns is obtained for $C_{4-h}$~=~10\,pF and $R_{4-h}$~=~1\,k$\Omega$. However, the corresponding amplitude decreases to 0.44\,$\cdot$\,$A_{ref}$. Similar results were also achieved with the 2\,$\times$\,2-hybrid configuration at 4.7\,pF \big($t_{rise}$~=~3.1\,ns and $A$~=~0.5\,$\cdot$\,$A_{ref}$\big). Since all amplitudes of the 4-hybrid connection are smaller than those of the 2\,$\times$\,2-hybrid arrays and also the smallest rise times were achieved with the 2\,$\times$\,2-hybrid configuration, a 4-hybrid connection was not considered further.

\begin{table}
\centering
\caption{Amplitude and rise time for the 4-hybrid connections.}
    \begin{tabular}{m{0.08\textwidth} m{0.085\textwidth} m{0.115\textwidth} m{0.1\textwidth}}
    \toprule[1pt]
    R$_{4-h}$ ($\Omega$) & C$_{4-h}$ (F) & $\nicefrac{A_{4-h}}{A_{ref}}$ & $t_{rise}$ (ns) \\ 
    \toprule[1pt]
    1k 			& 10 p  & 0.44 $\pm$ 0.09 		  & 3.26 $\pm$ 0.72   \\
    1k 			& 1 n   & 0.66 $\pm$ 0.12 		  & 4.79 $\pm$ 0.79   \\
    47 			& 1 n   & 0.84 $\pm$ 0.18 		  & 5.22 $\pm$ 0.96   \\
    47 			& 100 n & 0.90 $\pm$ 0.18 		  & 7.28 $\pm$ 1.46   \\
    \toprule[1pt]
    \end{tabular}
\label{Tab.2}
\end{table}

\subsection{Final 2\,$\times$\,2-hybrid circuit}
\noindent In order to determine the proper component sizes, the resistances (R3 = R4) and the capacitor ($C_{3}$) from \cref{Fig.8} were varied from 50\,$\Omega$ to 10\,k$\Omega$, and  4.7\,pF to 300\,nF, respectively. For the resistor variation, a clear peak in amplitude was observed at $R_{2x2}$~=~1\,k$\Omega$. As expected, a steeper signal edge is achieved with smaller capacitors $C_{2x2}$ and $C_{out}$. In this case, $C_{2x2}$~=~270\,pF was chosen as a middle ground between low rise time ($t_{rise}$~=~(3.7~$\pm$~0.3)\,ns) and sufficient amplitude ($\nicefrac{A_{2x2}}{A_{ref}}$~=~2.64~$\pm$~0.17).

Measurements without preamplifier, using the DC coupled S-out were performed to directly compare the signals of the 2\,$\times$\,2-hybrid circuit with the ones of the commercially available 2\,$\times$\,2 onsemi array consisting of the same SiPM type (onsemi ArrayJ-60035-4P). All cathodes of the onsemi ArrayJ-60035-4P are connected, and a breakout board with parallel connection was used. Omitting the high-pass created previously by $C_{out}$ and the terminating impedance (50 $\Omega$) is the reason for a bigger $t_{rise}$ than in the measurements before.

Compared to the one of the onsemi ArrayJ-60035-4P, the rise time of the 2\,$\times$\,2-hybrid array is more than halved, and the amplitude increased by a factor of 1.7 (\cref{Tab.3}).

\begin{table}
\centering
\caption{Direct comparison of the 2\,$\times$\,2 onsemi ArrayJ-60035-4P with the 2\,$\times$\,2-hybrid array (1\,k$\Omega$, 270\,pF) without $C_{out}$ and without preamplifier.}
	\begin{tabular}{l l l l}
	\toprule[1pt]
	Array 	& $\nicefrac{A_{2x2}}{A_{ref}}$ & $t_{rise}$ (ns) & J (ns) \\
	\toprule[1pt]
	ArrayJ-60035-4P  & 0.19 $\pm$ 0.02  &  17.0 $\pm$ 0.6 & 89 $\pm$ 10 \\
	2\,$\times$\,2-hybrid  & 0.32 $\pm$ 0.02 & 7.7 $\pm$ 0.7   & 24 $\pm$ 3 \\
	\toprule[1pt]
	\end{tabular}
\label{Tab.3}
\end{table}

\section{Preamplifier} \label{Sec:preamplifier}
\subsection{Amplifier types}
\noindent To find an optimal preamplifier for our SiPM signal, different types of operational amplifier-based preamplifiers have been tested. Three common types for amplifying a SiPM signal are the transimpedance amplifier and the inverting and non-inverting voltage amplifier \cite{WangComparison}. These circuits are discussed in the following. For all types of amplifiers, a low input impedance and a large signal bandwidth are important. An adder circuit for summing up the photocurrents of the individual SiPMs has also been considered.

\subsubsection{Adder (Add)}
\noindent The output $U_a$ of the adder circuit follows 
\begin{align}
    U_a = -R_f \sum_i{\frac{U_{Ei}}{R_i}}
\end{align}
with $R_f$ as the feedback resistor, $U_{Ei}$ as the applied voltages, and $R_i$ as the input resistors.

\subsubsection{Transimpedance amplifier (TIA)} \label{sec:TIA}
\begin{figure}
\centering
\includegraphics[width=0.423\textwidth]{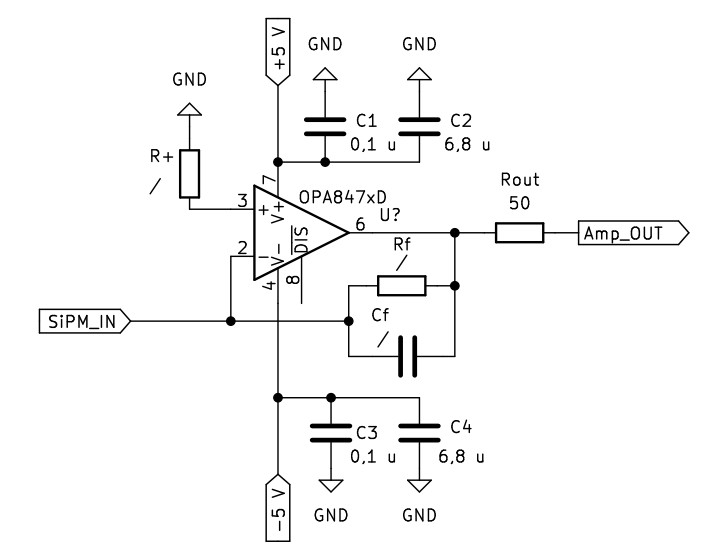}
\caption{Circuit diagram of the transimpedance amplifier.}
\label{Fig.9}
\end{figure}

\noindent A transimpedance amplifier has the property of converting current into voltage so that the amplification factor is expressed in units of impedance. The non-inverting input is grounded in this case (\cref{Fig.9}). Ideally, the input impedance should be 0\,$\Omega$. In reality, however, this is not possible at high frequencies, so the transimpedance amplifier must be stabilized by a capacitor $C_f$ in parallel with the feedback resistor. Its size can be estimated via
\begin{align}
C_f = \sqrt{\frac{C_{D}}{2\pi \cdot GBP \cdot R_f}}
\end{align}
where $C_D$ is the capacitance of the SiPMs, $GBP$ is the abbreviation for Gain-Bandwidth Product, and $R_f$ is the feedback resistor. According to \cite{fastPreamp}, the transimpedance gain is greater than the gain of a non-inverting voltage amplifier and also results in lower noise.
 
\subsubsection{Non-inverting amplifier (NIA)}
\begin{figure}
\centering
\includegraphics[width=0.36\textwidth]{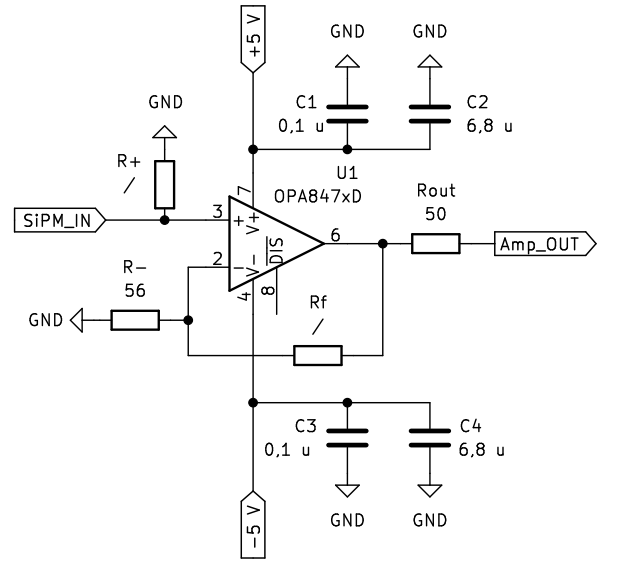}
\caption{Circuit diagram of the non-inverting voltage amplifier.}
\label{Fig.10}
\end{figure}

\noindent In the case of a voltage amplifier, the SiPM current must first be converted into a voltage pulse through a shunt resistor ($R_+$) connected to ground (\cref{Fig.10}). The voltage at the non-inverting input of the operational amplifier is then multiplied by the factor of the feedback resistor $R_f$. It is thus a non-inverting preamplifier whose amplification factor $G_+$ is defined by 
\begin{align}
    G_+ = 1+ \frac{R_f}{R_-}
\end{align}
where $R_{-}$ is the resistor connected to the inverting input and ground.

\subsubsection{Inverting amplifier (IA)}
\noindent For the commonly used inverting amplifier, the following applies:
\begin{align}
    U_a = - \frac{R_f}{R_{in}}U_E
    \label{FormelIV}
\end{align}
with $R_{in}$ as the input resistor and $U_E$ as the input voltage. 

\begin{figure}
\centering
\includegraphics[width=0.43\textwidth]{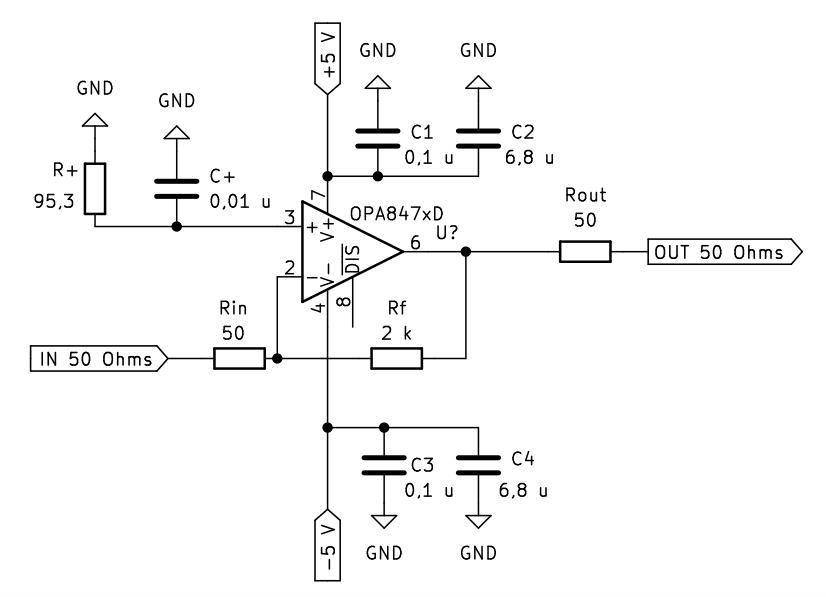}
\caption{Circuit diagram of the inverting amplifier from the OPA847 datasheet \cite{DatashOPA847}.}
\hfill\\[0.1cm]
\label{Fig.11}
\end{figure}

Compared to the non-inverting version, the inverting amplifier has the advantage of maintaining a constant noise gain $G_N$ with double signal amplification \cite{DatashOPA847}.
\begin{align}
    G_N = 1+\frac{R_f}{2\cdot R_{in}}
\end{align}
A proposed design of an inverting amplifier from the OPA847 datasheet is shown in \cref{Fig.11}.
Here, C1 to C4 are the buffer capacitors for the supply voltage, and in this case, a capacitor $C_{+}$ is used at the non-inverting input for compensation.

\subsection{Comparison of different amplifiers}
\noindent For all measurements, the high-frequency operational amplifier OPA847 was used, and the circuits were implemented on a PCB board with the variable resistors and capacitors being mounted as non-SMD components. All common rules for high-frequency designs, such as short signal paths and placement of components close to the IC, have to be implemented to ensure signal stability. As suggested in the datasheet of OPA847, the supply voltage was buffered with 0.1\,$\mathrm{\mu F}$ (ceramic) and 6.8\,$\mathrm{\mu F}$ (electrolytic) in parallel in order to get a high bandwidth decoupling. The input resistors for all circuits are set to $R_{in}$~=~50\,$\Omega$ because this value matches the termination impedance of the signal.

\cref{Tab.4} shows the amplification factors (AF) of the amplifiers (all used with $R_f$~=~4.7\,k$\Omega$, $C_{out}$~=~22\,pF and the 2\,$\times$\,2-hybrid SiPM array ($R_{2x2}$~=~1\,k$\Omega$ and $C_{2x2}$~=~270\,pF)). ``Add (4C)'' denotes a version of the adder circuit in which the signals from each SiPM anode were first AC coupled via $C_{out}$ and then added. Since this resulted in a relatively high rise time due to parallel capacitors, another version ``Add (1C)'' was tried. Here the photocurrents were first added and then decoupled with only one capacitor $C_{out}$~=~22\,pF.

\begin{table}
\centering
\caption{Comparison of different amplifier types based on the rise time ($t_{rise}$) and gain factor (AF) of the amplifier in combination with $C_{out}$ (22\,pF). $R_f$ was set to 4.7\,k$\Omega$ and the 2\,$\times$\,2-hybrid array (270\,pF, 1\,k$\Omega$) was used for photodetection.}
	\begin{tabular}{l l l l}
	\toprule[1pt]
	Amplifier type	& AF  & $t_{rise}$ (ns) & J (ns)  \\
	\toprule[1pt]
	Add (4C)  & 78 $\pm$ 12 & 7.3 $\pm$ 0.5   & 1.5 $\pm$ 0.4 \\
	Add (1C)  & 25 $\pm$ 4  & 6.1 $\pm$ 0.4   & 3.9 $\pm$ 0.7 \\
	TIA (stabilized)  & 31 $\pm$ 6   & 5.4 $\pm$ 0.5   & 2.9 $\pm$ 0.5 \\
	NIA	 & 16 $\pm$ 2  & 6.7 $\pm$ 1.5   & 6.7 $\pm$ 0.6 \\
	IA	& 42 $\pm$ 3 & 5.3 $\pm$ 0.3   & 2.04 $\pm$ 0.17\\
	\toprule[1pt]
	\end{tabular}
	\hfill\\[0.1cm]
\label{Tab.4}
\end{table}

The amplitude of the adder circuit is proportional to the number of SiPMs, so this circuit is ideal for a large number of SiPMs and gaining large amplitudes. However, the rise time of more than 6\,ns in the case of four connected SiPMs does not suit our needs.

Using $R_{f}$~=~4.7\,$k\Omega$ the transimpedance design had to be compensated. This results in an amplitude that is more than halved, compared to the uncompensated circuit with $R_f$~=~1.2\,k$\Omega$.

The non-inverting voltage amplifier was built according to the schematic in \cref{Fig.10} with $R_{+}$~=~56\,$\Omega$. Its gain is the lowest with 16. Combined with the knowledge that the inverting amplification shows a better signal-to-noise ratio (\cref{sec:TIA}), the non-inverting voltage amplifier was no longer considered.

In summary, the adder circuit and the transimpedance amplifier without $C_{f}$ are ideal for achieving gains up to a factor of 78. However, this results in a loss of rise time. Furthermore, the transimpedance amplifier was also very susceptible to oscillations. The non-inverting voltage amplifier is not convincing in terms of amplitude or rise time. So it confirms the statement of having a lower gain than the transimpedance amplifier mentioned in \cref{sec:TIA}. Best performing in terms of rise time was the inverting voltage amplifier. Due to a still quite high amplification factor of 42, the J-value is only lower for the 4C summing circuit, which is attributed to its very large amplitude. Additionally, the inverting amplifier offers a wide range of balancing amplitude and rise time by the choice of $R_f$. This variability is advantageous as the final time resolution can only be determined with the additional readout electronics and software.

\subsection{Dimensioning of the inverting amplifier} \label{Sec:dimensioningIA}
\noindent The final component values were found experimentally by varying them around reasonable starting values. The results show that the amplitude does not depend significantly on $R_+$. However, this is different for $t_{rise}$, which has a clear minimum at $R_+$~=~82\,$\Omega$. The capacitor given in the datasheet (\cref{Fig.11}) at the non-inverting input was not necessary to include, as the circuit is not susceptible to parasitic oscillations. 

By varying $R_{in}$, it can be observed that the rise time is virtually independent of this resistor since the variation occurs in the range below 50\,$\Omega$. Although the amplitude increases for smaller resistors, the tendency for output oscillation increases significantly. Therefore, the decision was to use a 50\,$\Omega$ input resistor, as suggested in the datasheet.

The amplitude and $t_{rise}$ increase with $R_f$. Only for $R_f$ below 1\,k$\Omega$ a susceptibility to oscillation could be observed. Hence, the choice of $R_f$ depends only on the compromise between amplitude and rise time. The lowest J between 2.7\,k$\Omega$ and 5.6\,k$\Omega$ is achieved with 4.7\,k$\Omega$. Therefore, this resistance was chosen.

\begin{table}[h]
\centering
\caption{Parameters of the final circuit.}
	\begin{tabular}{m{0.175\textwidth} m{0.165\textwidth} m{0.06\textwidth}}
	\toprule[1pt]
	Component & R ($\Omega$) & C (F)  \\
	\toprule[1pt]
	$R_f$	  & 4.7 k 		& - 	\\
	$R_+$	  & 82 			& - 	\\
	$R_{in}$  & 50 			& - 	\\
	$R_{2x2}$ & 1 k 			& - 	\\
	$C_{2x2}$ & - 			& 270 p \\
	$C_{out}$ & - 			& 15 p  \\
	\toprule[1pt]
	\end{tabular}
\label{Tab.5}
\end{table}

To determine the gain \textit{AF} of the amplifier, recorded pulses of the SiPMs were used as input, and together with the parameters listed in \cref{Tab.5}, it then resulted  in
\begin{align}
AF = 42 \pm 3 .
\end{align}
The signal rise time is increased by a factor of 
\begin{align}
\frac{t_{rise,in}}{t_{rise,out}} = 1.3 \pm 0.1
\end{align}
by the preamplifier circuit within these measurements. Using this inverting amplifier in combination with the 2\,$\times$\,2-hybrid array and $C_{out}$~=~15\,pF leads to a rise time of ($3.7\pm 0.3$)\,ns and an amplitude of (174.6 $\pm$ 0.3)\,mV at our setup.

\section{Conclusion} \label{Sec:outlook}
\noindent The purpose of this work was to develop the most suitable and cost-effective 2\,$\times$\,2 SiPM arrays and preamplifiers for a PET-like detector system intended to track densely packed and moving granular assemblies. Therefore, common SiPM interconnections (serial, parallel, and hybrid) were tested for their time performance.

It turned out that a hybrid interconnection where 2 SiPMs are serially connected in a 2-stage hybrid configuration offers the best results in terms of signal amplitude and rise time.

The corresponding optimal amplifier was selected by testing several types of operational amplifier-based preamplifiers like transimpedance amplifiers, inverting and non-inverting voltage amplifiers, and an adder circuit.

Considering all common rules for high-frequency designs, the inverting voltage amplifier resulted in the lowest achievable rise times while still maintaining sufficiently high gain. With a gain of 42, the signal rise time is affected by a factor of 1.3 by the preamplifier circuit.

\section*{Declaration of competing interest}
\noindent The authors declare that they have no known competing financial interests or personal relationships that could have appeared to influence the work reported in this paper.

\section*{Acknowledgments}
\noindent Funded by the Deutsche Forschungsgemeinschaft (DFG, German Research Foundation) -- Project-ID 422037413 -- TRR 287.

\bibliographystyle{elsarticle-num}
\bibliography{PaperBib}

\begin{thebibliography}{10}
\expandafter\ifx\csname url\endcsname\relax
  \def\url#1{\texttt{#1}}\fi
\expandafter\ifx\csname urlprefix\endcsname\relax\def\urlprefix{URL }\fi
\expandafter\ifx\csname href\endcsname\relax
  \def\href#1#2{#2} \def\path#1{#1}\fi

\bibitem{positronRangeNa-22}
L.~J{\o}dal, C.~Le~Loirec, C.~Champion, Positron range in pet imaging:
  non-conventional isotopes, Physics in Medicine \& Biology 59~(23) (2014)
  7419.

\bibitem{PEPTDevelopment1}
M.~Hawkesworth, C.~Bemrose, P.~Fowles, M.~O'Dwyer, Industrial application of
  positron emission tomography, Tomography and Scatter Imaging', N McCuaig and
  R Holt (Eds.), IOP Publishing, Bristol (1989) 67--79.

\bibitem{PEPTDevelopment2}
M.~Hawkesworth, D.~Parker, P.~Fowles, J.~Crilly, N.~Jefferies, G.~Jonkers,
  Nonmedical applications of a positron camera, Nuclear Instruments and Methods
  in Physics Research Section A: Accelerators, Spectrometers, Detectors and
  Associated Equipment 310~(1-2) (1991) 423--434.

\bibitem{PEPTReview}
C.~R.~K. Windows-Yule, M.~T. Herald, A.~L. Nicu{\c{s}}an, C.~S. Wiggins,
  G.~Pratx, S.~Manger, A.~E. Odo, T.~Leadbeater, J.~Pellico, R.~T.~M.
  de~Rosales, A.~Renaud, I.~Govender, L.~B. Carasik, A.~E. Ruggles,
  T.~Kokalova-Wheldon, J.~P.~K. Seville, D.~J. Parker,
  \href{https://doi.org/10.1088/1361-6633/ac3c4c}{Recent advances in positron
  emission particle tracking: a comparative review}, Reports on Progress in
  Physics 85~(1) (2022) 016101.
\newblock \href {https://doi.org/10.1088/1361-6633/ac3c4c}
  {\path{doi:10.1088/1361-6633/ac3c4c}}.
\newline\urlprefix\url{https://doi.org/10.1088/1361-6633/ac3c4c}

\bibitem{AnstiegszeitSzinti}
R.~Ogawara, M.~Ishikawa, Signal pulse emulation for scintillation detectors
  using geant4 monte carlo with light tracking simulation, Review of Scientific
  Instruments 87~(7) (2016) 075114.

\bibitem{SaintGobainSzintillator}
{Saint-Gobain Crystals},
  \href{https://www.crystals.saint-gobain.com/sites/hps-mac3-cma-crystals/files/2021-12/Organics-Plastic-Scintillators.pdf}{Organic
  scintillation materials and assemblies}, latest access: June 01, 2023 (2021).
\newline\urlprefix\url{https://www.crystals.saint-gobain.com/sites/hps-mac3-cma-crystals/files/2021-12/Organics-Plastic-Scintillators.pdf}

\bibitem{oppotsch2023simulation}
J.~Oppotsch, A.~Athanassiadis, M.~Fritsch, F.-H. Heinsius, T.~Held, N.~Hilse,
  V.~Scherer, M.~Steinke, U.~Wiedner,
  \href{https://www.sciencedirect.com/science/article/pii/S167420012300072X}{A
  simulation study for a cost-effective pet-like detector system intended to
  track particles in granular assemblies}, Particuology 84 (2024) 117--125.
\newblock \href {https://doi.org/https://doi.org/10.1016/j.partic.2023.03.005}
  {\path{doi:https://doi.org/10.1016/j.partic.2023.03.005}}.
\newline\urlprefix\url{https://www.sciencedirect.com/science/article/pii/S167420012300072X}

\bibitem{J-PET_medicalPerspectives}
P.~Moskal, E.~St{\k{e}}pie{\'n}, Prospects and clinical perspectives of
  total-body pet imaging using plastic scintillators, PET clinics 15~(4) (2020)
  439--452.

\bibitem{J-PET_PositroniumImaging}
P.~Moskal, K.~Dulski, N.~Chug, C.~Curceanu, E.~Czerwi{\'n}ski, M.~Dadgar,
  J.~Gajewski, A.~Gajos, G.~Grudzie{\'n}, B.~C. Hiesmayr, et~al., Positronium
  imaging with the novel multiphoton pet scanner, Science Advances 7~(42)
  (2021) eabh4394.

\bibitem{J-PET_Setup}
S.~Nied{\'z}wiecki, P.~Bia{\l}as, C.~Curceanu, E.~Czerwi{\'n}ski, K.~Dulski,
  A.~Gajos, B.~G{\l}owacz, M.~Gorgol, B.~Hiesmayr, B.~Jasi{\'n}ska, et~al.,
  J-pet: a new technology for the whole-body pet imaging, arXiv preprint
  arXiv:1710.11369 (2017).

\bibitem{BiasingReadout}
{Semiconductor Components Industries},
  \href{https://www.onsemi.com/pub/Collateral/AND9782-D.PDF}{Biasing and
  Readout of ON Semiconductor SiPM Sensors}, latest access: February 21, 2023
  (2019).
\newline\urlprefix\url{https://www.onsemi.com/pub/Collateral/AND9782-D.PDF}

\bibitem{Cryo}
W.~Ootani,
  \href{https://indico.gsi.de/event/6990/contributions/31545/attachments/22643/28404/ICASiPM_OverviewCryoReadout_Ootani.pdf}{Overview
  of readout techniques for cryogenic sipms}, latest access: February 21, 2023
  (2018).
\newline\urlprefix\url{https://indico.gsi.de/event/6990/contributions/31545/attachments/22643/28404/ICASiPM_OverviewCryoReadout_Ootani.pdf}

\bibitem{SiPMinNuclearPhysics}
F.~Simon, Silicon photomultipliers in particle and nuclear physics, Nuclear
  Inst. and Methods in Physics Research, A 926 (2019) 85--100.
\newblock \href {https://doi.org/10.1016/j.nima.2018.11.042}
  {\path{doi:10.1016/j.nima.2018.11.042}}.

\bibitem{SiPMDatash}
{Semiconductor Components Industries},
  \href{https://www.onsemi.com/pdf/datasheet/microj-series-d.pdf}{Datasheet
  J-Series SiPM Sensors}, latest access: December 12, 2022 (2021).
\newline\urlprefix\url{https://www.onsemi.com/pdf/datasheet/microj-series-d.pdf}

\bibitem{WangComparison}
M.~Wang, Y.~Wang, Q.~Cao, L.~Wang, K.~Jie, Y.~Xiao, Comparison of three
  pre-amplifier circuits for time readout of sipm in tof-pet detectors, 2019
  IEEE International Symposium on Circuits and Systems (ISCAS) (2019)
  Conference Date: 26--29 May 2019\href
  {https://doi.org/10.1109/ISCAS.2019.8702411}
  {\path{doi:10.1109/ISCAS.2019.8702411}}.

\bibitem{fastPreamp}
J.~Huizenga, S.~Seifert, F.~Schreuder, H.~van Dam, P.~Dendooven, H.~L{\"o}hner,
  R.~Vinke, D.~Schaart, A fast preamplifier concept for sipm-based
  time-of-flight pet detectors, Nuclear Instruments and Methods in Physics
  Research A 695 (2012) 379--384.
\newblock \href {https://doi.org/10.1016/j.nima.2011.11.012}
  {\path{doi:10.1016/j.nima.2011.11.012}}.

\bibitem{DatashOPA847}
{Texas Instruments},
  \href{https://www.ti.com/lit/ds/symlink/opa847.pdf}{Wideband, Ultra-Low
  Noise, Voltage-Feedback OPERATIONAL AMPLIFIER with Shutdown (Datasheet
  OPA847)}, latest access: November 11, 2022 (2008).
\newline\urlprefix\url{https://www.ti.com/lit/ds/symlink/opa847.pdf}

\end{thebibliography}

\end{document}